\documentclass[aps,pra,amsfonts,amssymb,epsfig,twocolumn,superscriptaddress,showpacs]{revtex4}

\usepackage{graphics,graphicx}
\usepackage{amsmath,bbm}
\usepackage{amstext}
\usepackage{amssymb}

\def\beq{\begin{equation}}
\def\eeq{\end{equation}}
\def\bea{\begin{eqnarray}}
\def\eea{\end{eqnarray}}
\providecommand{\openone}{\leavevmode\hbox{\small1\kern-3.8pt\normalsize1}}
\providecommand{\ket}[1]{|#1\rangle}
\providecommand{\oneket}[1]{\ket{1_{(#1)}}}
\providecommand{\auxket}[1]{\ket{1^\textrm{aux}_{(#1)}}}
\begin{document}

\title{High fidelity state transfer in binary tree spin networks}

\author{T. Tufarelli}
\affiliation{Scuola Normale Superiore, Piazza
dei Cavalieri 7, I-56126 Pisa, Italy}
\author{V. Giovannetti}
\affiliation {NEST CNR-INFM \& Scuola Normale Superiore, Piazza
dei Cavalieri 7, I-56126 Pisa, Italy}

\date{\today}

\begin{abstract}
Quantum state propagation over binary tree configurations
is studied in the context of quantum spin networks. 
For binary tree of order two a simple protocol is presented which allows to achieve arbitrary
high transfer fidelity. 
It 
does not require fine tuning of local fields and two-nodes coupling of the intermediate spins. Instead it assumes simple  local operations on the intended receiving node: their role  is to brake the transverse 
symmetry of the network that induces an effective refocusing of the
propagating signals. Some ideas on how to scale up these
effect to binary tree of arbitrary order are discussed.
\end{abstract}

\pacs{03.67.Hk, 03.67.Lx }

\maketitle

\section{Introduction}
The paradigmatic approach to quantum communication
assumes the possibility of  ``loading'' quantum information (i.e. qubits)
into mobile physical systems which are then transmitted 
from the sender of the messages
to their intended receiver. 
Such {\em flying qubit} architecture for quantum communication
 has found its natural implementation in optics where photons  play the role of information carriers.  In many respect this appears to be the most reasonable choice, specially when  long distance are involved in the communication. However   the recent development of controllable quantum many-body systems such as optical lattices~\cite{OL}, phonons in ion traps~\cite{IT}, Josephson arrays~\cite{JA}, and
polaritons in optical cavities~\cite{POC}, 
makes it plausible to consider  alternative  quantum communication scenarios
 such as the so 
called {\em quantum wire} architectures~\cite{BOSEREV}.
Here the transfer of quantum information proceeds over an extended network 
of coupled quantum systems (e.g. spins) which are {\em at rest} with respect to the communicating parties. In this case the messages are encoded into the internal states of the spins while the information flow proceeds  by their mutual  interactions   which, when properly tuned, induce a net transfer of  messages from
 two separate regions of the network~\cite{BOSE,LLOYD,SUB}. 
 The quantum wire architecture is of course of limited application, since it assumes the sender and the receiver to have access to the same quantum network 
(in any real implementation the latter will always have a reduced size).
However these techniques may play an important role in the creation of 
  clusters of otherwise independent  quantum computational devices. Furthermore
  the study of quantum network communication protocols is
  an ideal playground to test and device new quantum communication
  protocols.

Perfect transfer among any two regions
of a quantum network can always be achieved if one allows 
the communicating parties to have direct access 
on the individual nodes of the network
 (for instance this can be done by swapping sequentially 
the information from one node
to subsequent one). These strategies are however extremely demanding in terms of control
and, even in the absence of external noise, are arguably prone to error 
due to the large number of quantum gates that have to be applied to the system.
A less demanding approach consists in fixing the interaction of the network once for all 
and letting the Hamiltonian  evolution of the system to convey the sender message
to the receiver. In this context perfect transmission 
can be achieved either by engineering the spins couplings~\cite{key-11,NIKO,YUNG,YUNG1}, 
or by choosing proper encoding and decoding protocols~\cite{OSBORNE,HASEL,ENDG,BB,VGB}.

 \begin{figure}[t!]
\begin{center}
\includegraphics[scale=0.4]{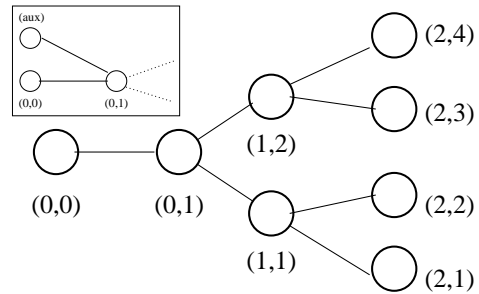}
 \caption{Spin tree of second order.  The nodes are connected through  edges which describe the ${XY}$ interactions 
defined by the Hamiltonian~(\ref{HAM}), and are 
 identified by a double index $(a,b)$,  with $(0,0)$ corresponding to the leftmost graph element. Inset:
 Auxiliary spin added to the BT in order to homogenise the effective couplings in the
block form representation~(\ref{hammat}). The additional
spin provide also a controllable trigger to start the information transfer. 
} \label{fig1}
\end{center}
\end{figure}

In this paper we discuss the propagation of quantum information over Bifurcation Tree (BT) quantum 
networks.  
Together with the star configuration the BT 
 configuration is arguably the most significant  network topology
 in circuit design. The former are typically  used as hubs to wire
 different computational devices (for an analysis of such system
 in the context of spin network communication see
 Ref.~\cite{YUNG}). Star configurations have been also extensively studied 
for entanglement distribution~\cite{HUTTON} and cloning~\cite{CLONE}.
 BT networks instead are employed to route 
  toward  external memory elements (i.e. database). 
The information  flow on unmodulated and uncontrolled BT was first discussed in Ref.~\cite{FARHI} while,
more recently, BT quantum networks have been employed to design efficient
quantum Random Access Memory elements~\cite{QRAM}. 

The paper is organised as follows. In Sec.~\ref{sec1}
we start analysing the first
 non trivial BT system
 introducing the notation and setting the problem.
In Sec.~\ref{sec2} we then describe a transfer protocol that allows
one to deliver a generic quantum message to any desired final
edges of the second order BT network by exploiting simple end
gates operations. In Sec.~\ref{sec3} we discuss various techniques
that allow us to scale up the protocol adapting it to BT of arbitrary order.
The paper finally ends with the conclusions and discussion in 
Sec.~\ref{sec4}.

\section{System description}\label{sec1}
First order BT networks are just  particular
instances  of  star networks~\cite{YUNG}. Consequently the 
simplest nontrivial examples of BT networks is
the  second order one shown in Fig.~\ref{fig1}. In the following we will assume 
the lines connecting the nodes to represent ${XY}$ (exchange) spin interactions (the results
however can be  generalised to include $XXZ$ or Heisenberg couplings).
The resulting Hamiltonian  is thus
\begin{equation}\label{HAM}
 H=\tfrac{J_{0}}{2}\sum_{<i,j>}(\sigma_{x}^{i}\sigma_{x}^{j}+\sigma_{y}^{i}\sigma_{y}^{j})+\sum_{j}\tfrac{\omega_{j}}{2}(\sigma_{z}^{j}+1)\;,
\end{equation}
where the summation is performed over all couples $i,j$ which are connected through an edge, where $\sigma_{x,y,z}^{i}$ are the Pauli matrices
associated with the $i$-th node, and where 
 the $\omega_i$'s appear in consequence of the interaction with local magnetic fields
 (in this expression the label $i$ an $j$ stands for the joint indexes $(a,b)$ of Fig.~\ref{fig1}).
 As usual~\cite{BOSEREV} we assume that  initially the system is in the ferromagnetic 
``all spin down''
ground state $|\O \rangle \equiv |0
\cdots 0\rangle$.
At time $t=0$ we then place an (unknown) qubit state $|\psi\rangle =\alpha|0\rangle+\beta|1\rangle$ on the left-most site $(0,0)$ (for instance by swapping it from an external
memory). With this choice the global state of the network is now
described by the vector
\begin{eqnarray}\label{ini}
|\Psi_{\text{in}} \rangle = 
\alpha |\O \rangle + \beta |1_{(0,0)}\rangle\;,
\end{eqnarray}
where $|1_{(a,b)}\rangle$ represents the network state where the node  $(a,b)$ is in the spin up state $|1\rangle$ while
the remaining ones are in the down state $|0\rangle$, i.e.
\begin{eqnarray} \label{vectors}
|1_{(a,b)}\rangle\equiv|0 \cdots 0\;  1_{(a,b)} \; 0 \cdots 0\rangle \;.
\end{eqnarray}
Knowing that the $z$ component of the total spin is preserved by the Hamiltonian evolution of the system  (i.e. $[H,S_z^{\textrm{tot}}]=0$) we can conclude that  the dynamics is costrained in the subspace of single-flips:
On this subspace $H$ acts in a very simple way that can be inferred from the graphical structure of the network, i.e. 
\begin{eqnarray}
H|1_{(a,b)}\rangle=\omega_{(a,b)}|1_{(a,b)}\rangle+J_0\sum_{(c,d)} |1_{(c,d)}\rangle \;,
\end{eqnarray}
where the sum is taken over all sites ${(c,d)}$ connected with $(a,b)$.
Our goal is to find a procedure that would allow us to transfer the qubit state $|\psi\rangle$ to one the right-most sites $(2,b)$ with $b\in \{ 1, 2,3, 4\}$ of our choice, i.e. 
\begin{eqnarray}\label{fini}
 |\Psi_{\text{in}}\rangle &\longrightarrow& |\Psi_{\text{fin}}^{(b)} \rangle \equiv  \alpha |\O \rangle + \beta |1_{(2,b)}\rangle\;.
\end{eqnarray}
Following Refs.~\cite{key-11,YUNG} one could try  to solve  this problem by fine tuning the 
parameters $J_0$ and $\omega_j$ of $H$ in  such a way that the free Hamiltonian evolution of the system
will be able to transform $|\Psi_{\text{in}} \rangle$ into $|\Psi_{\text{fin}}^{(b)} \rangle$ after some time interval $\tau$~\cite{NOTA1}.
This however  is  in general a quite complex calculation which entails to solve an inverse eigenvalue problem. 
Moreover, if any, the solutions obtained using such strategy will be arguably highly asymmetrical in the distribution of the local magnetic fields 
$\omega_j$'s. 
To avoid all this, here we will pursue a different  approach by limiting the freedom one has in choosing the Hamiltonian parameters but,
as in Refs.~\cite{HASEL,ENDG,VGB},
allowing local manipulation on the receiving node of the network (i.e.  $(2,b)$).
 Under these conditions we can show that a simple protocol exists that  realises
  the transformation~(\ref{vectors}) with arbitrary accuracy. It  assumes an homogeneous network structure where
  all the ratios $\omega_j/J_0$  are chosen to be  
 identical and equal to some fix value, and it is composed by the following three-steps: 
 \begin{itemize}
\item[{1.}]   the system is allowed to evolve freely under the action of $H$ for some time $\tau$; 
 \item[{2.}]   at this point on the receiving node $(2,b)$ is performed a fast (ideally instantaneous) 
  local transformation $\text{PS}_{(2,b)}$;
  \item[{3.}]   the network is then let evolve for an extra time interval $2 \tau$. 
\end{itemize}
 During the first step, due to the homogeneity of the Hamiltonian, the information flows along
  the left-right axis of the network while delocalising  along the south-north axis.  
  The value of  $\tau$ is approximatively the time interval an excitation takes   to travel  from the leftmost
 node $(0,0)$ to the rightmost  column formed by $(2.1)$, $(2,2)$, $(2,3)$ and $(2,4)$.
 The role of the local transformation $\text{PS}_{(2,b)}$ of step two is to brake the
  south-north symmetry   of the resulting  state  by flipping  the sign of a specific wave vector
    component. The system is then let evolve freely for a time interval which twice the initial one: 
    this is approximatively the time it takes an excitation to 
   leave the rightmost column, "bounce back" the leftmost node, and return to the rightmost network column.
  Due to the symmetry brake introduced at the second step, however, the signal will now  
   not diffuse over all the four sites  $(2.1)$, $(2,2)$, $(2,3)$ and $(2,4)$, but instead it will focus on 
   the intended receiving node $(2,b)$.  A detailed description of the protocol will be presented in Sec.~\ref{sec2}.
\subsection{Diagonalisation of the Hamiltonian}
To solve our problem we can exploit the fact that the ground state $|\O\rangle$ of the network  does not evolve to restrict ourselves to the case $\alpha =0$, i.e. $|\psi\rangle = |1\rangle$. 
We then simplify the structure of the Hamiltonian $(1)$ assuming  all $\omega_{j}$s to be identical, i.e. 
$\omega_j\equiv\omega$. In dealing with magnetic spins this means we are applying an homogeneous magnetic field of constant strength $B\propto-\omega$ all over the system. We could set $\omega=0$, since the energy is defined up to a constant, but we let it be nonzero 
to guarantee that $|\O\rangle$  is the ground state of the system.
 We now choose the following basis for the single excitation 
sector, that divides the Hamiltonian in invariant blocks:
\begin{eqnarray}\label{basefacile}
&&\text{B$_1$}\left\lbrace\begin{array}{l}
| v_0\rangle\equiv|1_{(0,0)}\rangle\\
| v_1\rangle\equiv|1_{(0,1)}\rangle\\ 
| v_2\rangle\equiv \tfrac{1}{\sqrt{2}} (|1_{(1,1)}\rangle+|1_{(1,2)}\rangle)\\ 
| v_3\rangle\equiv\tfrac{1}{2}(|1_{(2,1)}\rangle+|1_{(2,2)}\rangle+|1_{(2,3)}\rangle+|1_{(2,4)}\rangle)
\end{array}\right. \nonumber \\ 
&&\text{B$_2$}\left\lbrace\begin{array}{l}
| v_4\rangle\equiv\frac{1}{\sqrt{2}}(|1_{(1,1)}\rangle-|1_{(1,2)}\rangle)\\ | v_5\rangle\equiv\frac{1}{2}(|1_{(2,1)}\rangle+|1_{(2,2)}\rangle-|1_{(2,3)}\rangle-|1_{(2,4)}\rangle)
\end{array}\right. \nonumber \\ 
&&\text{B$_3$}\left\lbrace\begin{array}{l}
| v_6\rangle\equiv\frac{1}{2}(|1_{(2,1)}\rangle-|1_{(2,2)}\rangle+|1_{(2,3)}\rangle-|1_{(2,4)}\rangle)\\
| v_7\rangle\equiv\frac{1}{2}(|1_{(2,1)}\rangle-|1_{(2,2)}\rangle-|1_{(2,3)}\rangle+|1_{(2,4)}\rangle)\;.
\end{array}\right. \nonumber 
\end{eqnarray}
 In this basis the matrix representing $H$ is given by
\begin{equation} \label{hammat}
\left(\begin{array}{ccc}
\left(\begin{array}{cccc}
\omega&J_0\\
J_0&\omega&J\\
&J&\omega&J\\
&&J&\omega\\
\end{array}\right)\\
&\left(\begin{array}{cc}
\omega&J\\
J&\omega
\end{array}\right)\\
&&\left(\begin{array}{cc}
\omega&0\\
0&\omega
\end{array}\right)
\end{array}\right)
\end{equation}
with $J\equiv\sqrt{2}J_0$. This shows that  the evolution of the network can be 
effectively described as three independent linear chains, the first composed by $4$ nodes, and the 
other by $2$ elements each. The basic idea in deriving the above basis is that any state in the form $\ket{1_{(a,b)}}$
is decoupled from the ''singlet'' superposition $\frac{1}{\sqrt2}(\ket{10}-\ket{01})$ of the two nearest neighbours qubits on its right. E.g. The two states $\frac{1}{\sqrt2}(\oneket{2,b}-\oneket{2,b+1})$ with $b=1,3$ are decoupled from the whole network and provide an alternative basis for the block $B_3$.\\
Of special interest for us is of course the  block $\text{B}_1$ which is 
the only one to have an overlap with the input state~(\ref{ini}). 
It is clear that the case $J_0=J$ would be much simpler to deal with
(in this case for instance one could  adapt the linear chain analysis  of Refs.~\cite{key-11,BOSE} to
simplify the calculation).
Such option however is not possible
if we assume the coupling strengths of the network to be fixed a priori.
Anyway we can use a trick to obtain the same result without adjusting the coupling strength
which, as discussed in the final paragraph of the present section, allows us
to improve also the controllability of the setup.
Suppose in fact to modify the system by adding an additional  spin connected only to site $(0,1)$ with the usual ${XY}$ coupling of strength $J_0$, as shown in the inset of Fig.~\ref{fig1}.
With this choice 
the Hamiltonian~(\ref{HAM}) 
 is replaced by 
\begin{equation}
H_{\textrm{new}}\equiv H+\tfrac{J_{0}}{2}(\sigma_{\textrm{aux}}^{x}\sigma_{0,1}^{x}+\sigma_{\textrm{aux}}^{y}\sigma_{0,1}^{y})+\tfrac{\omega}{2}(\sigma^{z}_{\textrm{aux}}+1). \label{hnew} 
\end{equation}
Now that we enlarged the Hilbert space we have to deal with the 9-dimensional space of single flips.
However  since
the singlet state $\frac{1}{\sqrt{2}}(|1_{(0,0)}\rangle-|1_{\textrm{aux}}\rangle)$ is decoupled from the rest, 
if we encode the ``logic''  state $|1\rangle$ on the sending end of the network as 
$| v_0^{\textrm{new}}\rangle\equiv\frac{1}{\sqrt{2}}(|1_{(0,0)}\rangle+|1_{\textrm{aux}}\rangle)$ instead of using $|1_{(0,0)}\rangle$, not only we recover a dynamics costrained in an 8-dimensional space, but we obtain also an effective coupling of strength $J=\sqrt{2}J_0$ between $| v_0^{\textrm{new}}\rangle$ and $| v_1\rangle$ (here $|1_\text{aux}\rangle$
is the analogous of the states~(\ref{vectors}) with the spin up 
localised on the auxiliary node).
 The four dimensional block of  our effective Hamiltonian thus becomes:
\begin{equation} \label{hami4}
H_{(4)}\equiv\left(\begin{array}{cccc}
\omega&J\\
J&\omega&J\\
&J&\omega&J\\
&&J&\omega\\
\end{array}\right)\;.
\end{equation}
Following Ref.~\cite{key-11} this can be easily put in diagonal form obtaining the eigenvalues 
\begin{eqnarray}
E_{1} \equiv \tfrac{2\omega-(\sqrt{5}+1) J}{2}\;, 
\qquad  
E_{2} \equiv \tfrac{2\omega-(\sqrt{5}-1) J}{2}\;,
\nonumber \\
E_{3} \equiv \tfrac{2\omega+(\sqrt{5}-1) J}{2}\;, 
\qquad  \label{eigenv}
E_{4} \equiv \tfrac{2\omega+(\sqrt{5}+1) J}{2}\;,
\end{eqnarray}
with the corresponding eigenstates described by the vectors 
\begin{align}
&\ket{e_1}\equiv\tfrac{1}{\sqrt{5+\sqrt{5}}}\left(-1,\tfrac{1+\sqrt{5}}{2}
,-\tfrac{1+\sqrt{5}}{2},1\right),\nonumber\\
&\ket{e_2}\equiv\tfrac{1}{\sqrt{5-\sqrt{5}}}\left(1,\tfrac{1-\sqrt{5}}{2},\tfrac{1-\sqrt{5}}{2},1\right),\nonumber\\
&\ket{e_3}\equiv\tfrac{1}{\sqrt{5-\sqrt{5}}}\left(-1,\tfrac{1-\sqrt{5}}{2},-\tfrac{1-\sqrt{5}}{2},1\right),\nonumber\\
&\ket{e_4}\equiv\tfrac{1}{\sqrt{5+\sqrt{5}}}\left(1,\tfrac{1+\sqrt{5}}{2},\tfrac{1+\sqrt{5}}{2},1\right)\;,
\end{align}
(expressed in the basis $|v_0^{\text{new}}\rangle$, $|v_1\rangle$, $|v_2\rangle$ and $|v_3\rangle$).

A part from simplifying the spectral properties of the Hamiltonian, the introduction of the site {aux} adds also 
an additional feature that substantially  enhances  our ability of controlling the system.
We have already mentioned
 the singlet state $\frac{1}{\sqrt{2}}(|1_{(0,0)}\rangle-|1_{\textrm{aux}}\rangle)$ is decoupled from all others vectors of the system. Therefore 
we can ``entrap" our qubit of information at the leftmost end of the network for as much time as we like by encoding its 
logic $1$ component in  such singlet state.
When we want the transfer to start we simply apply a  local Phase shift $\textrm{PS}_\textrm{aux}$  on the auxiliary spin
that induces
the mapping $|1_{\textrm{aux}}\rangle\rightarrow-|1_{\textrm{aux}}\rangle$. This will  transform $\frac{1}{\sqrt{2}}(|1_{(0,0)}\rangle-|1_{\textrm{aux}}\rangle)$ 
into $|v_0^{\text{new}}\rangle$  bringing the encoded message into the four-dimensional
subspace associated with the Hamiltonian~(\ref{hami4}) and 
allowing the first step of the above protocol to begin. 
\section{Transfer protocol}\label{sec2}
In this section we analyse in detail the performance of the protocol defined in Sec.~\ref{sec1}.
Without loss of generality we consider the case in which the receiving node is $(2,1)$, i.e.  $b=1$. 
We recall  that our aim is to obtain the transition $| v_0^{\text{new}}\rangle\rightarrow|1_{(2,1)}\rangle$, and we notice that
\begin{eqnarray}
|1_{(2,1)}\rangle=\tfrac{1}{2}(| v_3\rangle+| v_5\rangle+| v_6\rangle+| v_7\rangle)
\label{finale} \;.
\end{eqnarray}
\paragraph*{The protocol: step {1.}}
In the first stage of the protocol the system is initialised  into $|v_0\rangle$ and freely evolves for some time $\tau$. The goal here is to find $\omega$ and $\tau$ such that this vector is mapped into $|v_3\rangle$,
which represents a symmetric combination in which the input
excitation is spread all over the rightmost nodes of the network.
As already noticed, this process is formally equivalent to the information transfer along a linear chain of $4$ spins coupled by uniform $XY$ first neighbours interactions. From Ref.~\cite{key-11} we know that such transferring cannot be exact. 
Nevertheless  the transfer fidelity can be made arbitrarily close to one.
Indeed defining $U(\tau) = \exp[ -i H\tau]$ and using Eq.~(\ref{eigenv}) we have 
\begin{eqnarray}
&&\langle v_3| U(\tau)| v_0^{\text{new}}\rangle = 
 \sum_{k=1}^4 e^{- i E_k t} \langle v_3 | e_k \rangle  \langle 
e_k | v_1\rangle 
\label{exp1} \\
&&= \nonumber
\tfrac{\sqrt{5} -5}{20} ( e^{-i E_1\tau} - e^{-i E_4\tau})
+ \tfrac{\sqrt{5} + 5}{20} ( e^{-i E_2\tau} - e^{-i E_3\tau})\;.
\end{eqnarray}
This  will be exactly one if one could find $\tau$ such that 
$e^{-i E_1\tau} = e^{-i E_3\tau} = -1$ and 
$e^{-i E_2\tau} = e^{-i E_4\tau} = 1$. Even though these conditions are impossible
to be satisfied exactly~\cite{key-11} an approximate solution is obtained
by choosing 
\begin{eqnarray}
\omega = \tfrac{7 + \sqrt{5}}{2} \;J = 
\tfrac{7 + \sqrt{5}}{\sqrt{2}} \; J_0
\label{imp1}\;,
\end{eqnarray}
and 
\begin{eqnarray}
\tau= \tau_n \equiv \frac{2n +1}{J} \pi \;,
 \end{eqnarray}
 with $n$ integer.
 Under this assumption Eq.~(\ref{exp1}) yields
 \begin{equation} \label{cos}
\langle v_3| U(\tau_n)| v_0^{\text{new}}\rangle =
\tfrac{1}{2} (1+e^{-i\varphi_n})\;,
\end{equation}
where $\varphi_n\equiv\sqrt{5}(2n+1)\pi$.
The exponential in Eq.~(\ref{cos}) 
never takes the value $1$ but, since $\sqrt{5}$ is an irrational number, approaches it indefinitely. Therefore for any $\varepsilon>0$ we can choose $n$ such that 
$|\langle v_3| U(\tau_n)| v_0^{\text{new}}\rangle| 
 \geqslant 1-\varepsilon$.
 As a result the state of the system, with high accuracy, is now described by the vector $|v_3\rangle$.
\paragraph*{The protocol: step {2.}}
As a second step we act locally on the node $(2,1)$, applying the phase shift  unitary transformation $\text{PS}_{21}$ which 
changes the sign to the state $|1_{(2,1)}\rangle$,
i.e.
\begin{eqnarray}
\textrm{PS}_{21} |1_{(2,1)}\rangle  = -|1_{(2,1)}\rangle \;,
\end{eqnarray}
while preserving the remaining single excitation states. This can be done, for instance, by acting with an intense magnetic field which
acts locally on $(2,1)$ for a time interval shorter than the characteristic times of the Hamiltonian $H$.
When acting on $|v_3\rangle$ 
the unitary $\textrm{PS}_{21}$ yields the following 
transformation,
\begin{equation}\label{superp}
\textrm{PS}_{21}| v_3\rangle=\frac{1}{2}(| v_3\rangle-| v_5\rangle-| v_6\rangle-| v_7\rangle)\;.
\end{equation}
This superposition contains the four states that compose the state $|1_{(2,1)}\rangle$, but the relative phases are wrong  -- see Eq.~(\ref{finale}). Luckily third step fixes this issue with free evolution only.
\paragraph*{The protocol: step {3.}}
Finally we just have to wait for a time $2\tau_n$ and the relative phases adjust themselves to give the state $|1_{(2,1)}\rangle$. In fact by explicit calculations it can be shown that
\begin{eqnarray}
\begin{array}{l}
\langle v_3| U(2\tau_n)| v_3\rangle=\frac{1}{2}(1+e^{-i2\varphi_n})\simeq1\;,\\
\\
\langle v_k| U(2\tau_n)| v_k\rangle=-e^{-i\varphi_n}\simeq-1\;,\qquad \mbox{for $k=5, 6, 7$.}
\end{array}
\end{eqnarray}
Both expressions are justified by the fact that $e^{-i\varphi_n}\simeq1$, and they imply that after third step we have reached state $|1_{(2,1)}\rangle$ with as good approximation as we like.
\newline

\begin{figure}[t!]
\begin{center}
\includegraphics[scale=0.4]{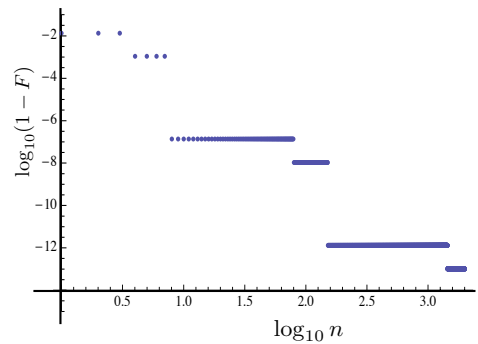}
\caption{Minimum "infidelity" $(1-F_n)$ achievable  according  Eq.~(\ref{fidelitynew}) 
for $n\in [0, N]$ as a function of $N$, i.e. 
$(1-F)_{\text{min}} \equiv \min_{n\leqslant N} (1 -F_{n})$. Notice that already for 
$n=8$ we get values of $F$ greater  than $1- 10^{-6}$. From the plot we can infer 
an almost linear dependence of $(1-F)_{\text{min}}$ with respect to  $N$, yielding  the following (approximated) behaviour $F_{\text{min}}\sim {1- c N^{-\gamma}}$
with $\gamma\sim 1$ and $c$ being  constant.} \label{fig4}
\end{center}
\end{figure}
The above operations can be summarised by the application of the unitary operator 
\begin{eqnarray}
V_n\equiv U(2\tau_n)\; \text{PS}_{(2,1)} \; 
U(\tau_n)\label{vn}\;.
\end{eqnarray}
 Therefore the resulting transfer fidelity can be expressed as
\begin{eqnarray} \label{fidelitynew}
F_n&\equiv& |\langle1_{(2,1)}| V_n|v_0^{\text{new}}\rangle|^2 
= \left| \tfrac{ 1+ e^{-i\varphi_n}}{8}\right|^2 
\nonumber \\
&&\times  \nonumber 
\left| 3e^{-i\varphi_n} + \tfrac{ 1  + e^{-2i\varphi_n}}{2}
+ (1  + e^{-i\varphi_n})^2 \right|^2 \\
 &&  = \tfrac{1}{4} \cos^2(\varphi_n/2) [ 3 -\cos^2(\varphi_n/2) ]^2 \;,
\end{eqnarray}
which  approaches $1$ if $e^{-i\varphi_n}\simeq1$
giving us the desired result -- see Fig.~\ref{fig4}.

\section{Some ideas to scale up the system}\label{sec3}
\begin{figure}[t!]
\begin{center}
\includegraphics[scale=1]{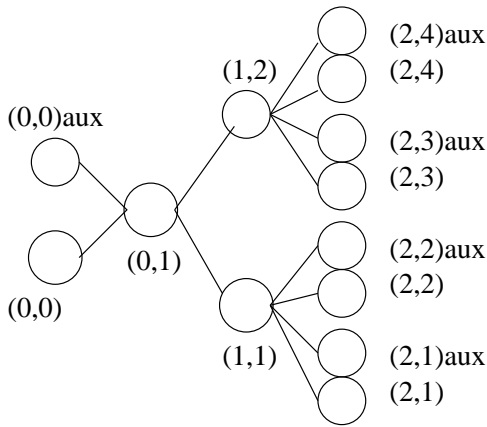}
\caption{Modified tree for the purpose of scaling up. Each rightmost qubit is accompanied now by an auxiliary, moreover all the rightmost couplings are adjusted by a factor ${1}/{\sqrt2}$.} \label{fig3}
\end{center}
\end{figure}

Unfortunately our protocol doesn't extend easily to higher-order trees (or at least we couldn't find a simple way of doing it).
The idea we pursued in trying to scale up a second order BT is to connect in some way the ends of a tree to the beginning of another. The resulting structure isn't anymore a tree of the type described above, but it's still a valid mean to obtain a larger number of outputs. As an example we could connect 
(say) two second order trees to the ends of a first order tree to obtain a  $8$-outputs quantum switch, or four second order trees to the
 ends of a fifth second order tree to obtain a $16$-outputs quantum switch -- see Fig.~\ref{fig3}.
The former setup can be solved by properly merging the protocol of Ref.~\cite{YUNG} with our second order BT propagation scheme:
this however will require to employ non-uniform magnetic fields at least for the first spins and does not admit simple concatenation.
We thus decided to focus on the second architecture which instead 
 can be trivially concatenated to form larger setup.
We found a relatively simple way to make the required connections, but at the expenses of considering some coupling strength engineering and including antiferromagnetic interactions, which means that the
 ``all down'' configuration is no longer the ground state, although still stationary.
A combination of time evolution and a phase shift will do all the work.
First of all each receiving end of a tree must be accompanied by an auxiliary qubit, as done before for the sending end. 
In analogy to what happened introducing $(0,0)_{\textrm{aux}}$, it is easy to see that the rightmost singlets \[\ket{s_{2b}}\equiv\frac{1}{\sqrt{2}}(|1_{(2,b)}\rangle-|1^{\textrm{aux}}_{(2,b)}\rangle)\;,\] are isolated, while the corresponding triplets \[\ket{t_{2b}}\equiv\frac{1}{\sqrt{2}}(|1_{(2,b)}\rangle+|1^{\textrm{aux}}_{(2,b)}\rangle)\;,\] interact with the network and evolve, with an effective coupling strength $\sqrt{2}$ times the original one. 
In order for our protocol to be still valid, we need to modify the coupling strengths of the rightmost branches so that matrix~(\ref{hami4}) (now with $J_0=J$) remains unchanged. Moreover the local operation on the receiving end $(2,b)$ must now be performed simultaneously on $(2,b)$ and $(2,b)_{\textrm{aux}}$, i.e. we must now apply $\text{PS}_{2b}+\text{PS}_{2b}^{\textrm{aux}}$.
In this way once the excitation reaches one of the end-triplets of the tree it can be trapped there with a Phase Shift on the auxiliary qubit of that site, storing information in the relative singlet. To clarify this, we outline that in this new configuration our protocol is capable of achieving the (approximate) transfer
\begin{equation}
\alpha |\O \rangle + \beta|1_{(0,0)}\rangle\rightarrow\alpha\ket{\O}+\beta\ket{t_{2b}}.
\end{equation}
Now by applying a local phase shift $\text{PS}_{2b}^{\textrm{aux}}$ the state is transformed into
\begin{equation}
\alpha\ket{\O}+\beta\ket{s_{2b}},
\end{equation}
which is decoupled from the rest.\\
If by some means we could transfer this state to the singlet at the beginning of the next tree (denoted by primed indexes), i.e. obtain the state
\begin{equation}
\alpha\ket{\O}+\beta\frac{1}{\sqrt2}(\oneket{0',0'}-\auxket{0',0'})\equiv\alpha\ket{\O}+\beta\ket{s_{0'0'}}\;,
\end{equation}
we could then perform the local operation $\textrm{PS}_{(0',0')}^\textrm{aux}$ to obtain the corresponding triplet state\[\alpha\ket{\O}+\beta\ket{t_{0'0'}},\] that can be transferred along the new tree with the usual protocol.
\begin{figure}[t!]
\begin{center}
\includegraphics[scale=1]{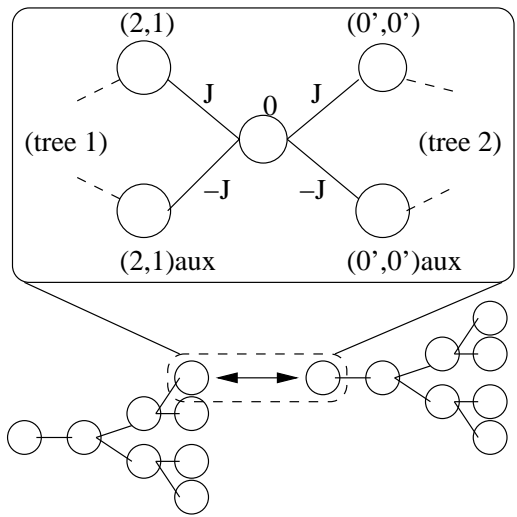}
\caption{The ``singlet link'' used to connect the outputs of the first tree to the inputs of the next column of trees (only the first one is represented here). The auxiliary qubits are not represented for simplicity.
} \label{fig5}
\end{center}
\end{figure}
The structure shown in Fig.~\ref{fig5} (that we will call  ``singlet link'') achieves perfect transfer between two singlets, since in the subspace $\{\ket{s_{21}},\ket{1_0},\ket{s_{0'0'}}\} $ it is equivalent to a chain of length 3 with constant couplings~\cite{key-11}, moreover the evolution of $\ket{t_{21}}$ is decoupled from that of $\ket{t_{0'0'}}$ thanks to the opposite signs of the couplings along the vertical axis.
The lines stand for ${XY}$ interaction of strength $J$ (ferromagnetic) and $-J$ (antiferromagnetic) respectively.
As an example we have considered site $(2,1)$ of a second-order tree plus its auxiliary qubit, connected with site $(0',0')$ plus its auxiliary of another tree.
We outline again that we are working in the subspace of single flips, as our Hamiltonian still conserves $S^{\textrm{tot}}_z$. In the considered example we have 
\begin{align*}
&H\ket{s_{21}}=\omega\ket{s_{21}}+\sqrt2J\ket{1_0},\\
&H\ket{t_{21}}=\omega\ket{t_{21}}+J\oneket{1,1},\\
&H\ket{s_{0'0'}}=\omega\ket{s_{0'0'}}+\sqrt2J\ket{1_0},\\
&H\ket{t_{0'0'}}=\omega\ket{t_{0'0'}}+\sqrt2\oneket{0',1'};
\end{align*}
where $\ket{t_{0'0'}}\equiv\frac{1}{\sqrt2}(\oneket{0',0'}+\auxket{0',0'})$.\\
We can see from the above equations that once the information enters a tree through a triplet state it doesn't come out of it until we make a Phase Shift on the desired end (and at the right time!). At this point the information goes to the singlet and propagates to the starting singlet of another tree, thanks to the singlet link, then it is transferred to the corresponding triplet with a local Phase Shift and propagation begins on the next tree. We shall repeat this procedure until information reaches the desired end on the last array of trees. Of course we must control a priori the total error due to the presence of second order trees, so to fix a ``single-tree time'' $\bar \tau$ which gives a satisfactory overall transfer fidelity.

\section{Conclusions}\label{sec4}
In this paper we have presented a protocol for quantum state transfer on BT spin networks
of order two. As in Ref.~\cite{HASEL,VGB,ENDG} it is based on the local operations which must be performed on the 
receiving nodes.  
Differently from~\cite{VGB,ENDG} however it does not involve swapping operation between the receiving nodes and external memories and  arbitrarily high fidelity can be 
obtained in just three operational steps. Generalisation of this techniques to higher orders BT is currently under investigation: arguably this will involve more complex 
ends gates operations possibly on more than one of the rightmost nodes.
We have however provided  a simple way to scale up the problem by concatenating 
smaller BT network through connecting gates which can be turned on and off by
simple local phase gate transformations.

We thank R. Fazio and D. Burgarth for comments and discussions.
This work was in part founded by the Quantum Information program of Centro Ennio De Giorgi of the Scuola Normale Superiore.


\begin{thebibliography}{99}
\bibitem{NOTA1}Along this line, for instance,
one could map the propagation over the network into a simpler problem by setting a subset of the  $\omega_j$s 
to a value $\Lambda$ much greater then the remaining constants of the systems. This will induce an effective
decoupling of the selected nodes from the remaining part of the network, simplifying the underline topology.
\end{thebibliography}
\end{document}